\documentclass[twocolumn,showpacs,preprintnumbers,amsmath,amssymb]{revtex4}

\usepackage{amssymb}
\usepackage{graphicx}
\usepackage{dcolumn}
\usepackage{bm}
\usepackage{color}

\begin{document}

\title{Field-induced avalanche to the ferromagnetic state in the
phase-separated ground state of manganites }
\author{F. M. Woodward$^1$}
\author{J. W. Lynn$^1$}
\author{M. B. Stone$^2$}
\author{R. Mahendiran$^2$}
\author{P. Schiffer$^2$}
\author{J. F. Mitchell$^{4}$}
\author{D. N. Argyriou$^{3}$}
\author{L. C. Chapon$^{5}$}
\affiliation{$^1$NIST Center for Neutron Research, National Institute of 
Standards and Technology, Gaithersburg, MD 20899-8562\\
$^2$Department of Physics and Materials Research Institute, Pennsylvania
State University, University Park, PA 16802\\
$^{4}$Materials Science Division, Argonne National Laboratory, Argonne, IL
60439\\
$^{3}$Hahn-Meitner Institute, Berlin D-14109, Germany
$^{5}$ISIS department, Rutherford Appleton Laboratory, Chilton Didcot, 
OX11 0QX,UK}
\date{\today}

\begin{abstract}
Perovskite manganite compounds such as 
Pr$_{1-x}$(Ca$_{1-y}$Sr$_{y}$)$_{x}$MnO$_{3}$ can be tuned to exhibit a 
metastable ground state where two magnetic/crystallographic phases coexist 
in zero magnetic field.  Field-dependent neutron
diffraction measurements on both poly- and single-crystal samples with a 
range of Pr, Ca, and Sr dopings(0.3$\le$x$\le$0.35 and y$\le$0.30)
reveal that the charge-ordered, antiferromagnetic phase of the ground state 
suddenly and irreversibly jumps to the ferromagnetic state.  The transition occurs 
spontaneously at some time after the field is set above a threshold field, 
indicating that once the transition is initiated an avalanche occurs that 
drives it to completion. 
\end{abstract}

\maketitle

Manganese oxide perovskites display a variety of complex and
interesting behavior resulting from the coupling of electronic, magnetic,
and structural degrees of freedom.  \cite{Goodenough55,Dagotto01} One such
phenomenon is the inhomogeneous coexistence of a charge and orbitally (CO) ordered
antiferromagnetic (AFM) state with a structurally distinct
ferromagnetic metallic phase (FMM) in a phase separated ground state.  
Experimental evidence\cite{Raveau95,Deac01,Blake02,Yaicle03} and computational results\cite
{Dagotto03} indicate the phases are arranged as FM domains embedded in a
CO/AFM matrix.  The delicate energy balance that exists between these
phases is easily tipped by external perturbations such as magnetic 
field\cite{Baca02,Yoshizawa95,Yaicle03}, electric field\cite{Stankiewicz00},
electron\cite{Hervieu99} or x-ray irradiation\cite{Cox98}, which destroy 
the CO/AFM phase and drive the system irreversibly into the FMM state.  

A model system for such phase-separated behavior is the hole-doped Pr$_{1-x}$
Ca$_{x}$MnO$_{3}$ (PCMO) system in the range x=0.3-0.45\cite{Jirak, Raveau97,Martin99},  
which can be fine-tuned by additional substitutions such
as Sr for Ca, Co for Mn, or Ga for Mn\cite{Yaicle03}.  An applied field
gradually converts the AFM phase to the FMM phase as expected, but recent
low temperature isothermal magnetization M(H) data for these PCMO compounds
reveal novel jumps in the magnetization that occur at discrete threshold
fields\cite{PCMO, Hebert02, Hardy03R,Hardy03}.  The rapid onset of these jumps at low
temperatures and their appearance in polycrystalline samples rule out the
usual phenomena such as spin-flop or metamagnetic transitions; such
transitions are only sharp for a narrow range of field directions with
respect to atomic spin directions in the crystal, and they do not exhibit a
sudden onset at low T (well below the magnetic ordering temperature)\cite{Carlin}.  To
investigate the microscopic origin of these jumps in the magnetization, we
have carried out field-dependent neutron diffraction and inelastic
measurements on both polycrystalline and single crystal samples.  We find
that at low T both the charge and antiferromagnetic order parameters
suddenly and irreversibly collapse above a threshold field, with 
a concomitant jump in the ferromagnetic order parameter.  
Our results indicate that there are two important factors leading to this unique
behavior.  One is the remarkable isotropy of the ferromagnetic
system, which makes the internal magnetic energy, to a very good approximation, 
independent of crystallographic direction.  The second factor
concerns the lattice strain\cite{Radaelli01,Hardy03m} that develops at the boundaries between 
the FM and CO/AFM phases.  This strain inhibits the smooth growth of FM domains,
resulting in a stick-slip growth of domains.  When one of the distorted CO/AFM
domains does transform to the FMM phase, the net magnetization in the vicinity of this
new FMM domain subsequently jumps, causing other CO/AFM regions to convert in an
avalanche\cite{Bak96,Hebert02} into a new phase fraction of CO-AFM/FMM domains. The process is
strongly time-dependent in that once the transformation is initiated, all
(or a large fraction) of the sample transforms spontaneously.  Once the FMM
state is established, returning the field to zero leaves the system in the
FMM phase, as there is no driving force to restore the CO/AFM state.  The
transformation then is completely irreversible, with the CO/AFM state only
being restored by warming above the ferromagnetic Curie temperature ($
\approx 100K$).

The samples investigated here were prepared by the floating zone technique
and have three compositions of varying Sr doping; a 2.7 g Pr$_{0.65}$(Ca$
_{0.75}$Sr$_{0.25}$)$_{0.35}$MnO$_{3}$ (PCSMO75) single crystal, a 3.6 g
textured polycrystalline sample and 5 g powder sample (for profile refinements) 
of Pr$_{0.65}$(Ca$_{0.70}$Sr$_{0.30}$)$_{0.35}$ MnO$
_{3}$ (PCSMO70), and a 1.4 g Pr$_{0.7}$Ca$_{0.3}$MnO$_{3}$ (PCMO) single
crystal.  Zero field neutron data were collected at the NIST Center for
Neutron Research on the BT7 triple axis spectrometer, and field dependent 
data on the BT2 and BT9 triple axis spectrometers in a 
vertical field superconducting magnet with a dilution insert.
Diffraction data were taken with a pyrolytic graphite (PG) monochromator set
to 14.7 meV, no analyzer, and coarse collimation (nominally 40$^{\prime }$
).  Inelastic measurements on the PCSMO75 single crystal were taken on BT9
using either a PG(002) or Ge(311) monochromator, and a PG(002) analyzer.  In
all cases, we employed a PG filter to suppress higher-order wavelength
contaminations.    Complete powder diffraction data were collected at the Hahn-Meitner
Institute at a series of fields (T = 1.5 K) with a vertical field magnet on the E9
spectrometer and a Ge (311) monochromator.
Small angle neutron scattering data were collected as a 
function of temperature and time on NG7 SANS for the single crystal sample of 
Pr$_{0.65}$(Ca$_{0.75}$Sr$_{0.25}$)$_{0.35}$MnO$_3$.  A wavelength of 
$\lambda$ = 10 {\AA} and a $\frac{\Delta\lambda}{\lambda}$ = 0.22 were employed for the 
measurements.  The sample was mounted in a closed cycle refrigerator equipped with 
Si windows.  Transport measurements were made using a four probe AC technique 
and magnetization data were collected using a commercial SQUID magnetometer.

\begin{figure}[tbp]
\includegraphics[width=2.75in]{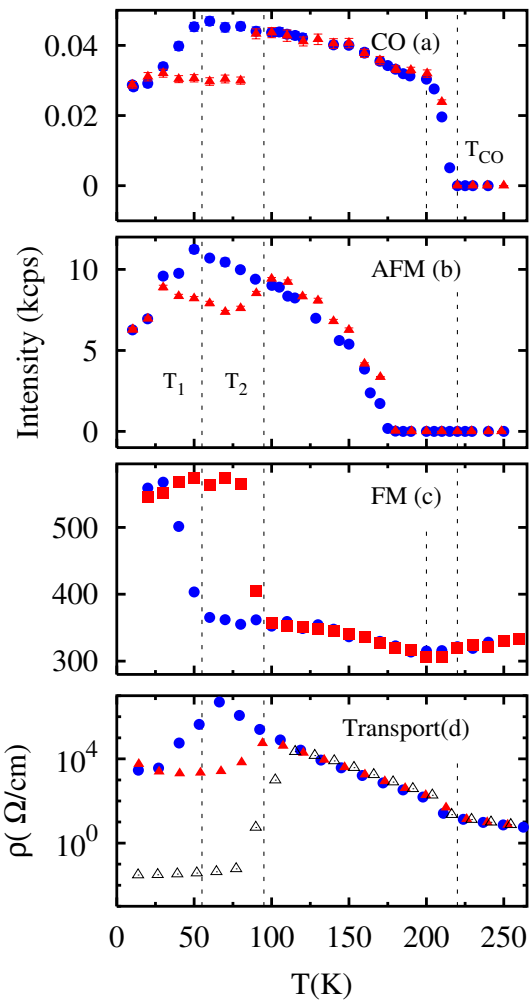}
\caption{(color online) Temperature dependent data comparing resistance 
and integrated intensity for the CO, AFM and FM peaks from a 
single crystal PCSMO75 sample.
Solid circles (\textcolor{blue}{\protect\large $\bullet $}) 
represent zero field data collected on cooling, solid triangles 
(\textcolor{red}{$\blacktriangle $}) are zero field data collected on
warming, open triangles ($\triangle $) represent zero field 
data collected upon warming after application of a 5 T field.  
(a) shows intensity of the (2$\frac{{1}}{{4}}$ 1$\frac{{3}}{{4}}$ 0) CO peak,
(b) ($\frac{{1}}{{2}}$ 0 0) antiferromagnetic peak, and 
(c) (1 0 0) ferromagnetic peak collected as a function of temperature. 
(d) Transport data showing resistance (semi-log scale) as a function of 
temperature.  The dashed vertical lines represent, from right to left, the CO 
transition, the Curie temperature, decrease in FM upon warming, rapid
increase in FM upon cooling.}
\label{ord}
\end{figure}

The competition between the CO/AFM and FM phases of PCSMO75 is evident in
the zero-field order parameters shown in Fig. \ref{ord}. Charge
ordering, which occurs below T$_{CO}$ = 220 K, results in a structural distortion 
producing new peaks at positions such as (2$\frac{1}{4}$ 1$\frac{3}{4}$ 0)
(Fig. \ref{ord}(a)).\cite{note1}  The distorted CO phase then orders 
antiferromagnetically at T$_{N}$ = 170 K (Fig. \ref{ord}(b)).  
Integrated intensity data at the (1 0 0) Bragg peak as a function of 
temperature reveal the onset of FM order at T$_c$ = 200 K, reflected in the CO
data as a slight change in slope.  Growth of the FM intensity, initially suppressed 
by the CO/AFM matrix\cite{Blake02},  begins to increase rapidly at lower temperatures, 
exhibiting the strong hysteresis seen in Fig. 1(c), with T$_1$= 55 K on cooling and 
T$_2$ = 95 K  on warming.  Changes in CO and AFM intensity are 
coincident with changes in the FM intensity at T$_1$ and T$_2$, 
demonstrating that the FM phase forms at the expense of the CO/AFM phase.

Changes in transport data, collected on a polycrystalline sample of PCSMO75,
shown in the semi-log plot of resistance as a function of temperature 
(Fig. \ref{ord}(d)), mirror the neutron results in Fig. \ref{ord}(a)-(c).  
The CO transition is evident in the transport data as a slight increase in the 
resistance at 210 K, for both warming and cooling measurements.  The resistance
reaches a maximum upon cooling at 65 K,  just before the temperature where the 
neutron data indicate a maximum in the CO phase.  The rapid increase of ferromagnetism 
at 55 K results in a decrease of the CO phase concomitant with the decrease in resistance.  
The fall in resistance is attributed to a larger portion of the sample entering the 
conducting FM phase.  Upon warming, the resistivity does not change until the FM 
phase falls below detectable levels in the neutron data, around 95 K.  Zero field
resistance data, collected in a field induced (5 T) FMM state, 
represented as $\triangle$ in Fig. \ref{ord}(d), indicate that the FMM state 
is eliminated at 110 K, just above the temperature where the FM signal disappears in 
the zero field data.

\begin{figure}[tbp]
\includegraphics[width=2.75in]{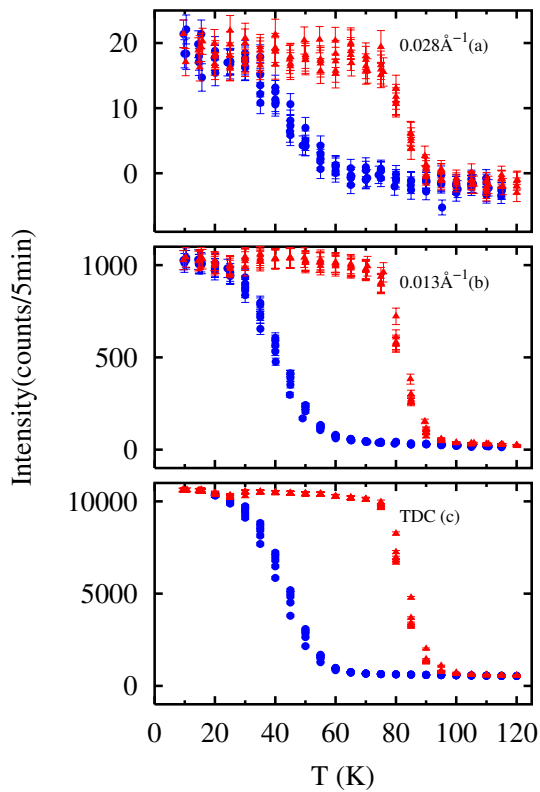}
\caption{(color online) SANS data collected on NG7 showing intensity plotted 
as a function of temperature.  Blue circles (\textcolor{blue}{\protect\large $\bullet$}) 
are data collected upon cooling, and red triangles (\textcolor{red}{$\blacktriangle$}) 
are data collected upon warming.  Fig. (a) and (b) show intensity changes at the specific 
q values 0.028 and 0.013 {\AA}$^{-1}$ respectively. (c) shows total detector counts, 
for the area detector.}
\label{pcsmo75-sans}
\end{figure}
The SANS data as a function of temperature sampled at two q values are 
shown in Fig. \ref{pcsmo75-sans}(a), q= 0.0278 {\AA}$^{-1}$ and 2(b) q=0.0127 {\AA}$^{-1}$.  
These data have been processed by subtracting a high temperature background, 
T= 320 K, as well as correcting for detector and instrumental configurations.  
For a ferromagnet typically there are two contributions in a SANS measurement.  
The small Q intensity originates from domains and domain walls and tracks a power 
of the magnetization, while the intensity at the larger Q originates from (dynamic) spin 
correlations\cite{Lynn98}.  Fig. \ref{pcsmo75-sans}(c) shows intensity integrated 
over the entire area of the detector (total detector counts) as a 
function of temperature.  The maximum changes in intensity occur for 
the temperatures T= 50 K upon cooling and T= 90 K upon warming, consistent with the 
hysteresis in the zero field order parameters.  The irreversibility demonstrates 
that the phase fraction population is metastable. The time dependence of the 
intensity was also checked by making five consecutive measurements at the same temperature.  
Over this 25 minute time span (five data sets at five minutes per point) the intensity 
drifts by a few percent, slowly increasing upon cooling and decreasing upon warming, 
indicating that the system has a long equilibration time.  This drift is 
most apparent in Fig. \ref{pcsmo75-sans}(a).  At higher q values 
(greater than ~0.03 {\AA}$^{-1}$) one typically expects to see critical scattering for
a second order ferromagnetic transition, but here the data only show the same sharp 
onset in intensity and hysteresis, indicating that 
ferromagnetism is already established in this temperature range.  
\begin{figure}[tbp]
\includegraphics[width=2.75in]{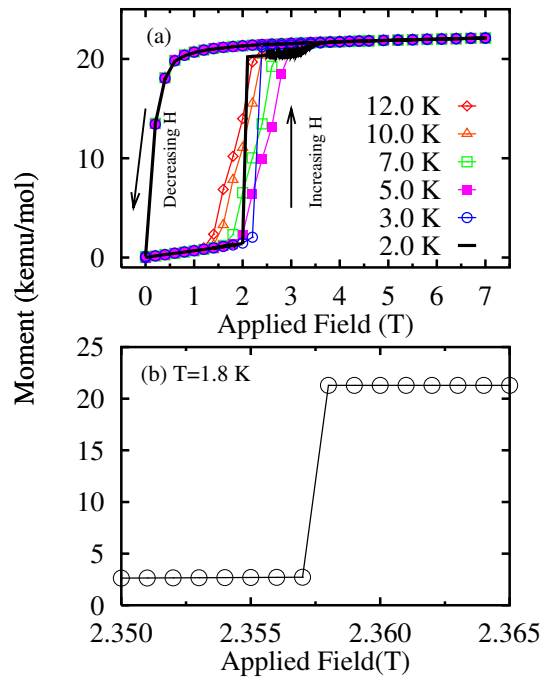}
\caption{(color online) Isothermal magnetization data for a
polycrystalline sample of PCSMO75 collected at several
temperatures.  (a) Shows the temperature dependence of the
transition.   The 2 K data (solid line) shows a clear step feature at
approximately H = 2 T, which becomes rounded and moves to lower magnetic 
field at elevated temperature.  A smaller step is also observed at 3.5 T (see text).
(b)  Sudden jump in the magnetization, indicating that any width of 
the step feature at T=1.8 K finds is less than 10 Oe.}
\label{mvh}
\end{figure}

\begin{figure}[tbp]
\includegraphics[width=2.75in]{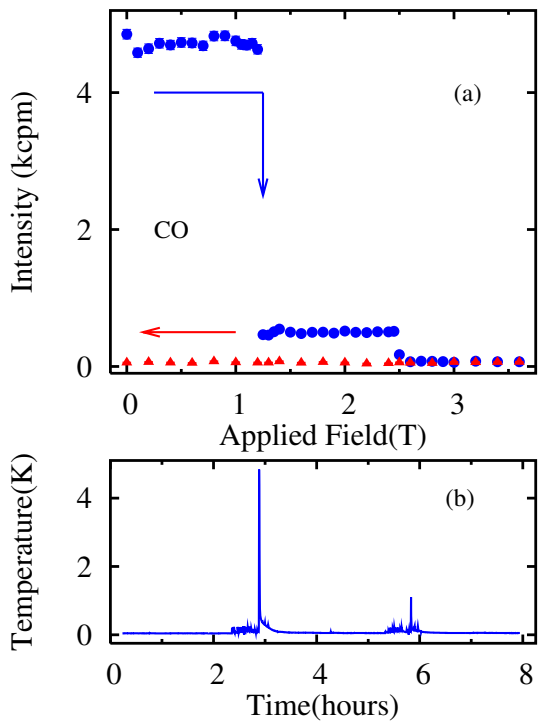}
\caption{(color online)(a) Neutron diffraction data for single crystal PCSMO75 
at T = 0.08 K, showing peak intensity as a function of applied field for the 
CO peak (2$\frac{{1}}{{4}}$ 1$\frac{{3}}{{4}}$ 0).  Solid circles 
(\textcolor{blue}{\protect\large $\bullet$}) are data collected while increasing 
the field from the zero field-cooled state.  Solid triangles 
(\textcolor{red}{$\blacktriangle$}) are data collected as the field was 
returned to zero.  (b) Sample temperature as a function of time 
during the PCSMO75 neutron diffraction measurements.  The jumps in sample temperature
versus time are coincident with the magnetization steps.}
\label{pcsmo75}
\end{figure}

Isothermal magnetization, M(H), curves for a zero field cooled 
polycrystalline sample of PCSMO75 are plotted in Fig. \ref{mvh}(a) for several 
temperatures.  Between successive measurements, the sample was warmed to 200 K, 
above the FM transition as observed in Fig \ref{ord}(c)-(d), to avoid 
thermal hysteresis effects in M(H).  
Below 2 K, a sharp step in M(H) is visible in Fig \ref{mvh}(a) at 1.8 T.  A second,
smaller stepped transition is observed at H = 3.5 T in M(H) and  both transitions 
are clearly present in the field dependent 
neutron data for PCSMO75, Fig \ref{pcsmo75}.  
A comparison of M(H) curves in Fig. \ref{mvh}(a) shows the temperature dependence of 
the magnetic transition.  As the temperature is elevated, the once sharp transition 
( for T$\le$2 ) shows a weak temperature dependence, becoming rounded and moving to 
somewhat smaller magnetic fields.  The data in Fig. \ref{mvh}(b) show the low
temperature transition to be quite sharp, less than 10 Oe.  The difference between the 
transition field between Fig. \ref{mvh}a and Fig. \ref{mvh}b is due to the size of the step 
used in approaching the transition.

Fig. \ref{ord} shows that at low temperatures the CO/AFM and FM
phases coexist in the ground state.  An applied magnetic field will change
the relative energetics of these two phases, and this behavior is clearly shown in 
Fig. \ref{pcsmo75}(a) for the intensity of the CO peak.  These data were 
obtained by zero-field cooling the sample to T = 80 mK.  The CO intensity 
is observed to change abruptly twice, first falling by an order of magnitude 
at 1.25 T, then vanishing at 2.5 T.  Both the field position and the size of these
steps are in good agreement with the magnetization data for this
sample as shown in Fig. \ref{mvh}.  The results clearly demonstrate that the step
increase in magnetization originates from a change in the phase fraction of
CO/AFM to FM phases, rather than a jump in the size of the (atomic) ordered
moment of the ferromagnet.  For this particular sample, we observe two
steps where the system is driven through a metastable FM/CO phase fraction
to the homogeneous FMM ground state by the applied field.  The
difference in critical fields between the neutron data, collected at 0.08 K 
and the magnetization data, collected at 1.8 K, is a result of the weak 
temperature dependence for the transition, previously reported
by Mahendiran et al\cite{PCMO} in the isothermal magnetization data.

One of the interesting aspects of these steps is that there is a large
release of energy when the transition occurs, as shown Fig. \ref{pcsmo75}
(b).  Here we plot the time dependence of the thermometer attached to the 
sample stage of the dilution refrigerator.  At the first step there is a 
large spike in the temperature when the sample transforms, from 0.08 K to 5.0 K, 
which subsequently returns to 0.08 K over a period of about ten minutes.  Similar behavior 
was observed at the second smaller step at 2.50 T, where the temperature 
reaches 1.4 K.  We estimate the magnetic energy gained,
$\Delta $E = \textbf{$\mu \cdot $B}, by increasing 
the ferromagnetic phase fraction for the 1.25 T transition 
(assuming roughly half the sample converts as suggested by the powder
diffraction fits for the PCSMO70 sample) is only $\sim $20\% of that
needed to raise the temperature by $\Delta $T = 4.9 K.  
It is therefore
clear that substantial additional heat, associated with the collapse of the
CO phase, must be released at this structural/magnetic transition.
This observation is consistent with a very recent report of 
the magnetocaloric effect in manganite samples.\cite{Ghivelder03}

\begin{figure}[tbp]
\includegraphics[width=2.75in]{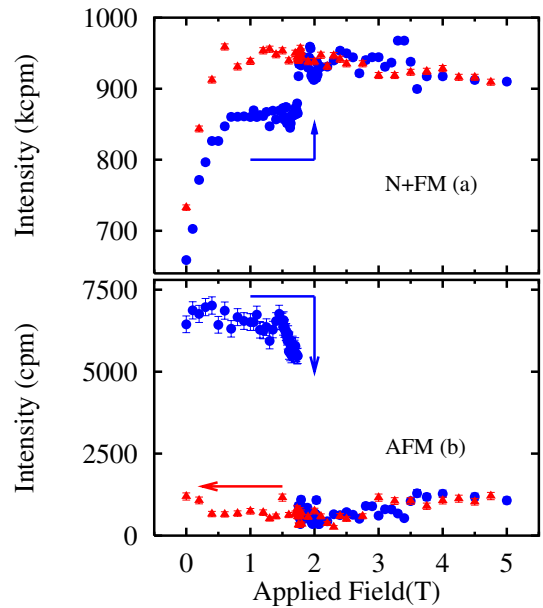}
\caption{(color online)Neutron data for polycrystalline PCSMO70 at T = 0.1 K
showing integrated intensity as a function of applied field.  (a) nuclear and ferromagnetic 
intensity at (1 0 0) and (b) antiferromagnetic signal at ($\frac{1}{2}$ 0 0).  Solid circles 
(\textcolor{blue}{\protect\large $\bullet $}) represent increasing the field
from a zero field cooled state.  Solid triangles (\textcolor{red}{$\blacktriangle $}) represent
returning the field to zero.}
\label{pcsmo70}
\end{figure}

Reducing the Sr doping by 5\%, from PCSMO75 to PCSMO70, results in a
compound exhibiting only one low temperature step transition in the
magnetization at 1.75 T.  High resolution powder diffraction data show that 
initially 65 \% of the sample is in the ferromagnetic phase, and this 
phase fraction jumps to 100\% between 1.1 and 1.9 T.  
A detailed field dependence from the polycrystalline sample
is shown in Fig. \ref{pcsmo70}.
We see (Fig. \ref{pcsmo70}(a)) that the FM intensity 
increases smoothly with field as the domains reorient, and reaches a plateau
around 0.75 T.  At H$_{c}$ = 1.75 T an 8\% jump is observed in the FM intensity.  
The small size of this jump is attributed to the strong extinction for
this peak, which has both a structural and ferromagnetic component.  At the
identical field the AFM intensity (Fig. \ref{pcsmo70}
(b)) abruptly vanishes.  The abrupt disappearance of the AFM intensity, combined 
with the sudden decrease of the CO intensity, demonstrates that the transition
originates from a change in phase fraction rather than spin canting.
The absence of an AFM signal in Fig. \ref{pcsmo70}
(b) when the field is returned to zero indicates the system is locked into the 
FMM state.

\begin{figure}[tbp]
\includegraphics[height=2.75in,angle=-90]{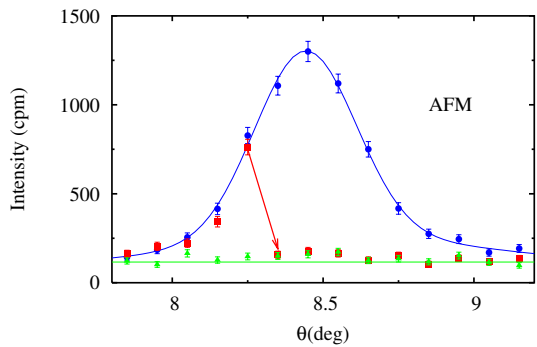}
\caption{(color online)Rocking curves for the antiferromagnetic Bragg peak [q= 
($\frac{1}{2}$ 0 0)] for the PCSMO70 sample collected at T= 2.0 K.  Data at 1.74 T (
\textcolor{blue}{\protect\large $\bullet $}) show a complete rocking curve.  The data for
1.75 T (\textcolor{red}{$\blacksquare $}) begin to follow the same curve, but suddenly the
intensity disappears.  The rocking curve at 1.76 T (\textcolor{green}{$\blacktriangle $}) shows
that the peak is completely gone.}
\label{collapse}
\end{figure}

The integrated intensity data shown in Fig. \ref{pcsmo70} were obtained by
rocking the sample through the Bragg reflection.  These data
reveal the process by which the sudden collapse of the CO/AFM phase
occurs.  Fig. \ref{collapse} shows three separate rocking curves for
PCSMO70 collected at the AFM peak (T = 2.0 K) for three values of applied 
magnetic field.  The data at 1.74 T show a normal rocking curve. The field 
was then increased to 1.75 T and another scan was initiated.  The rocking 
curve is essentially identical to the data at 1.74 T until the sixth point, 
when the intensity suddenly vanishes.  The collapse thus occurs spontaneously 
at some time after the field is set above 
the threshold, indicating that once the transition is initiated an avalanche 
occurs driving the transformation to completion.  The transition appears 
instantaneous on the time scale of seconds needed to collect the data and
explains the apparent perfect sharpness of the steps.

To compare the mixed-phase FM state with the field-induced FMM state 
we measured the ferromagnetic spin waves 
with inelastic neutron scattering.  In both cases the spin waves
are well described by a gapless, isotropic FM model; E$_{sw}=Dq^{2}$ where 
$D$ is the spin stiffness constant\cite{Lynn96}.  For the initial zero field-cooled FM
state we obtained $D$ = 126(6) meV-{\AA}$^2$ , while we measured $D$ = 134(5)
meV-{\AA}$^2$ for the FMM state.  These values are consistent with 
other observations of D in the induced metallic state\cite{Baca02}, in contrast to
the ferromagnetic insulating state where D is about $\frac{1}{3}$ of these 
values\cite{Dai01,Adams03}.  
The statistically insignificant difference in spin stiffness between states 
indicates that the zero-field phase-separated FM state is metallic and 
identical to the FMM state.

In the zero field phase-separated model of PCSMO, 
FM domains nucleate in a CO/AFM matrix\cite{Blake02}.  Strain develops 
at the interface between the P{\it 21/m}, CO/AFM phase, and the P{\it nma}, FM phase, 
suppressing the growth of FM domains.  An applied field aligns the randomly 
oriented FM clusters and lowers the energy  of the ferromagnetic state.
The spin wave data show the ferromagnet to be isotropic, 
so the energy of the FM state does not exhibit a
significant dependence on the crystallographic direction of the applied field, explaining 
the rather remarkable result that identical behavior is observed
for polycrystalline and single crystal samples.  We believe 
the underlying cause of the discontinuous behavior to be a result of the
lattice strain that develops at the interface between the CO/AFM and FM phases, which
blocks the simple percolative growth of the FM domains that cause the CO/AFM 
state to become metastable at low temperatures.  Indeed, simulations adjusted to 
model strain at the interface between the two phases successfully predict discontinuous 
steps in resistivity as a function of temperature, and such simulations may be used 
to model the stick-slip growth of the FM domains in these materials as a function of 
field.\cite{Dagotto03}  Removing the thermal fluctuations at low temperatures sharpens 
the magnetic interfaces, making it more difficult for the distorted CO/AFM phase to be 
converted to the FMM phase.  When the magnetic field overcomes this energy the CO/AFM 
domain can suddenly convert to the FMM phase.  The net local magnetization increase 
combined with the release of strain held in the lattice distortions causes a sudden 
change in phase fraction.  Collapse of the CO/AFM phase can be arrested at intermediate 
CO/AFM:FM phase fractions, resulting in multiple steps in the magnetism.  
This behavior also explains the effect of field cooling on 
the magnetization steps.  Cooling the sample in an applied field increases the initial 
size of the FM domains, removing the lower energy (lower field) transitions.  These 
FM domains have a larger surface area requiring greater fields to overcome the 
barrier to growth, increasing the transition field.

The collapse of the CO/AFM phase can be described quantitatively with avalanche theory 
if the collapse of the system is characterized by power law behavior in one or more 
parameters.\cite{Bak96}  In the present case the time scales are much too fast to
measure with scattering techniques, but may be accessible by measuring the 
magnetic (Barkhausen) noise spectrum.
The amplitude and frequency of the magnetic noise should be directly related
to the rate of the collapse of CO/AFM domains into the FMM state, and if the collapse 
is an avalanche process then the noise frequency spectrum could be examined for 
power law behavior.
It will be particularly interesting to determine whether the phase separated PCMO systems,
with appropriately tuned compositions, can be connected to other complex systems such
as vortex flow\cite{Field95} or sandpiles\cite{Bak96} which display avalanching behavior.

This research was performed while F. M. Woodward held a National Research
Council Research Associateship Award at the NIST Center for Neutron
Research.  This work was supported in part by the U.S. Department of Energy, 
Office of Science, under Contract No. W-31-109-ENG-38.  Work at Penn State was 
funded by NSF grant DMR-01-01318.

\bibliography{pcmo-prb}

\begin{thebibliography}{30}
\expandafter\ifx\csname natexlab\endcsname\relax\def\natexlab#1{#1}\fi
\expandafter\ifx\csname bibnamefont\endcsname\relax
  \def\bibnamefont#1{#1}\fi
\expandafter\ifx\csname bibfnamefont\endcsname\relax
  \def\bibfnamefont#1{#1}\fi
\expandafter\ifx\csname citenamefont\endcsname\relax
  \def\citenamefont#1{#1}\fi
\expandafter\ifx\csname url\endcsname\relax
  \def\url#1{\texttt{#1}}\fi
\expandafter\ifx\csname urlprefix\endcsname\relax\def\urlprefix{URL }\fi
\providecommand{\bibinfo}[2]{#2}
\providecommand{\eprint}[2][]{\url{#2}}

\bibitem[{\citenamefont{Goodenough}(1955)}]{Goodenough55}
\bibinfo{author}{\bibfnamefont{J.~B.} \bibnamefont{Goodenough}},
  \bibinfo{journal}{Phys. Rev.} \textbf{\bibinfo{volume}{100}},
  \bibinfo{pages}{564} (\bibinfo{year}{1955}).

\bibitem[{\citenamefont{Dagotto et~al.}(2001)\citenamefont{Dagotto, Hotta, and
  Moreo}}]{Dagotto01}
\bibinfo{author}{\bibfnamefont{E.}~\bibnamefont{Dagotto}},
  \bibinfo{author}{\bibfnamefont{T.}~\bibnamefont{Hotta}}, \bibnamefont{and}
  \bibinfo{author}{\bibfnamefont{A.}~\bibnamefont{Moreo}},
  \bibinfo{journal}{Phys. Rep.} \textbf{\bibinfo{volume}{344}},
  \bibinfo{pages}{1} (\bibinfo{year}{2001}).

\bibitem[{\citenamefont{Raveau et~al.}(1995)\citenamefont{Raveau, Maignan, and
  Caignaert}}]{Raveau95}
\bibinfo{author}{\bibfnamefont{B.}~\bibnamefont{Raveau}},
  \bibinfo{author}{\bibfnamefont{A.}~\bibnamefont{Maignan}}, \bibnamefont{and}
  \bibinfo{author}{\bibfnamefont{C.}~\bibnamefont{Caignaert}},
  \bibinfo{journal}{J. Solid State Chem.} \textbf{\bibinfo{volume}{117}},
  \bibinfo{pages}{424} (\bibinfo{year}{1995}).

\bibitem[{\citenamefont{Deac et~al.}(2001)\citenamefont{Deac, Mitchell, and
  Schiffer}}]{Deac01}
\bibinfo{author}{\bibfnamefont{I.~G.} \bibnamefont{Deac}},
  \bibinfo{author}{\bibfnamefont{J.~F.} \bibnamefont{Mitchell}},
  \bibnamefont{and} \bibinfo{author}{\bibfnamefont{P.}~\bibnamefont{Schiffer}},
  \bibinfo{journal}{Phys. Rev. B} \textbf{\bibinfo{volume}{63}},
  \bibinfo{pages}{172408} (\bibinfo{year}{2001}).

\bibitem[{\citenamefont{Blake et~al.}(2002)\citenamefont{Blake, Chapon,
  Radaelli, Argyriou, Gutmann, and Mitchell}}]{Blake02}
\bibinfo{author}{\bibfnamefont{G.~R.} \bibnamefont{Blake}},
  \bibinfo{author}{\bibfnamefont{L.}~\bibnamefont{Chapon}},
  \bibinfo{author}{\bibfnamefont{P.~G.} \bibnamefont{Radaelli}},
  \bibinfo{author}{\bibfnamefont{D.~N.} \bibnamefont{Argyriou}},
  \bibinfo{author}{\bibfnamefont{M.~J.} \bibnamefont{Gutmann}},
  \bibnamefont{and} \bibinfo{author}{\bibfnamefont{J.~F.}
  \bibnamefont{Mitchell}}, \bibinfo{journal}{Phys. Rev. B}
  \textbf{\bibinfo{volume}{66}}, \bibinfo{pages}{144412}
  (\bibinfo{year}{2002}).

\bibitem[{\citenamefont{Yaicle et~al.}(2003)\citenamefont{Yaicle, Martin,
  Jirak, Fauth, Andre, Suard, Maignan, Hardy, Retoux, Hervieu
  et~al.}}]{Yaicle03}
\bibinfo{author}{\bibfnamefont{C.}~\bibnamefont{Yaicle}},
  \bibinfo{author}{\bibfnamefont{C.}~\bibnamefont{Martin}},
  \bibinfo{author}{\bibfnamefont{Z.}~\bibnamefont{Jirak}},
  \bibinfo{author}{\bibfnamefont{F.}~\bibnamefont{Fauth}},
  \bibinfo{author}{\bibfnamefont{G.}~\bibnamefont{Andre}},
  \bibinfo{author}{\bibfnamefont{E.}~\bibnamefont{Suard}},
  \bibinfo{author}{\bibfnamefont{A.}~\bibnamefont{Maignan}},
  \bibinfo{author}{\bibfnamefont{V.}~\bibnamefont{Hardy}},
  \bibinfo{author}{\bibfnamefont{R.}~\bibnamefont{Retoux}},
  \bibinfo{author}{\bibfnamefont{M.}~\bibnamefont{Hervieu}},
  \bibnamefont{et~al.}, \bibinfo{journal}{Phys. Rev. B}
  \textbf{\bibinfo{volume}{68}}, \bibinfo{pages}{224412}
  (\bibinfo{year}{2003}).

\bibitem[{\citenamefont{Burgy et~al.}(2003)\citenamefont{Burgy, Dagotto, and
  Mayr}}]{Dagotto03}
\bibinfo{author}{\bibfnamefont{J.}~\bibnamefont{Burgy}},
  \bibinfo{author}{\bibfnamefont{E.}~\bibnamefont{Dagotto}}, \bibnamefont{and}
  \bibinfo{author}{\bibfnamefont{M.}~\bibnamefont{Mayr}},
  \bibinfo{journal}{Phys. Rev. B.} \textbf{\bibinfo{volume}{67}},
  \bibinfo{pages}{014410} (\bibinfo{year}{2003}).

\bibitem[{\citenamefont{Fernandez-Baca
  et~al.}(2002)\citenamefont{Fernandez-Baca, Dai, Kawano-Furukawa, Yoshizawa,
  PLummer, Katano, Tomioka, and Tokura}}]{Baca02}
\bibinfo{author}{\bibfnamefont{J.~A.} \bibnamefont{Fernandez-Baca}},
  \bibinfo{author}{\bibfnamefont{P.}~\bibnamefont{Dai}},
  \bibinfo{author}{\bibfnamefont{H.}~\bibnamefont{Kawano-Furukawa}},
  \bibinfo{author}{\bibfnamefont{H.}~\bibnamefont{Yoshizawa}},
  \bibinfo{author}{\bibfnamefont{E.~W.} \bibnamefont{PLummer}},
  \bibinfo{author}{\bibfnamefont{S.}~\bibnamefont{Katano}},
  \bibinfo{author}{\bibfnamefont{Y.}~\bibnamefont{Tomioka}}, \bibnamefont{and}
  \bibinfo{author}{\bibfnamefont{Y.}~\bibnamefont{Tokura}},
  \bibinfo{journal}{Phys. Rev. B} \textbf{\bibinfo{volume}{66}},
  \bibinfo{pages}{1} (\bibinfo{year}{2002}).

\bibitem[{\citenamefont{Yoshizawa et~al.}(1995)\citenamefont{Yoshizawa, Kawano,
  Tomioka, and Tokura}}]{Yoshizawa95}
\bibinfo{author}{\bibfnamefont{H.}~\bibnamefont{Yoshizawa}},
  \bibinfo{author}{\bibfnamefont{H.}~\bibnamefont{Kawano}},
  \bibinfo{author}{\bibfnamefont{Y.}~\bibnamefont{Tomioka}}, \bibnamefont{and}
  \bibinfo{author}{\bibfnamefont{K.}~\bibnamefont{Tokura}},
  \bibinfo{journal}{Phys. Rev. B} \textbf{\bibinfo{volume}{52}},
  \bibinfo{pages}{R13145} (\bibinfo{year}{1995}).

\bibitem[{\citenamefont{Stankiewicz et~al.}(2000)\citenamefont{Stankiewicz,
  Ses$\acute{e}$, Garc$\acute{i}$a, Blasco, and Rillo}}]{Stankiewicz00}
\bibinfo{author}{\bibfnamefont{J.}~\bibnamefont{Stankiewicz}},
  \bibinfo{author}{\bibfnamefont{J.}~\bibnamefont{Ses$\acute{e}$}},
  \bibinfo{author}{\bibfnamefont{J.}~\bibnamefont{Garc$\acute{i}$a}},
  \bibinfo{author}{\bibfnamefont{J.}~\bibnamefont{Blasco}}, \bibnamefont{and}
  \bibinfo{author}{\bibfnamefont{C.}~\bibnamefont{Rillo}},
  \bibinfo{journal}{Phys. Rev. B} \textbf{\bibinfo{volume}{61}},
  \bibinfo{pages}{11236} (\bibinfo{year}{2000}).

\bibitem[{\citenamefont{Hervieu et~al.}(1999)\citenamefont{Hervieu,
  Barnab$\acute{e}$, Martin, Maignan, and Raveau}}]{Hervieu99}
\bibinfo{author}{\bibfnamefont{M.}~\bibnamefont{Hervieu}},
  \bibinfo{author}{\bibfnamefont{A.}~\bibnamefont{Barnab$\acute{e}$}},
  \bibinfo{author}{\bibfnamefont{C.}~\bibnamefont{Martin}},
  \bibinfo{author}{\bibfnamefont{A.}~\bibnamefont{Maignan}}, \bibnamefont{and}
  \bibinfo{author}{\bibfnamefont{B.}~\bibnamefont{Raveau}},
  \bibinfo{journal}{Phys. Rev. B} \textbf{\bibinfo{volume}{60}},
  \bibinfo{pages}{R726} (\bibinfo{year}{1999}).

\bibitem[{\citenamefont{Cox et~al.}(1998)\citenamefont{Cox, Radaelli, Marezio,
  and Cheong}}]{Cox98}
\bibinfo{author}{\bibfnamefont{D.~E.} \bibnamefont{Cox}},
  \bibinfo{author}{\bibfnamefont{P.~G.} \bibnamefont{Radaelli}},
  \bibinfo{author}{\bibfnamefont{M.}~\bibnamefont{Marezio}}, \bibnamefont{and}
  \bibinfo{author}{\bibfnamefont{S.-W.} \bibnamefont{Cheong}},
  \bibinfo{journal}{Phys. Rev. B} \textbf{\bibinfo{volume}{57}},
  \bibinfo{pages}{3305} (\bibinfo{year}{1998}).

\bibitem[{\citenamefont{Jirak et~al.}(1985)\citenamefont{Jirak, Krupicka,
  Simsa, Dlouha, and Vratislav}}]{Jirak}
\bibinfo{author}{\bibfnamefont{Z.}~\bibnamefont{Jirak}},
  \bibinfo{author}{\bibfnamefont{S.}~\bibnamefont{Krupicka}},
  \bibinfo{author}{\bibfnamefont{Z.}~\bibnamefont{Simsa}},
  \bibinfo{author}{\bibfnamefont{M.}~\bibnamefont{Dlouha}}, \bibnamefont{and}
  \bibinfo{author}{\bibfnamefont{S.}~\bibnamefont{Vratislav}},
  \bibinfo{journal}{J. Magn. Magn. Mater} \textbf{\bibinfo{volume}{53}},
  \bibinfo{pages}{153} (\bibinfo{year}{1985}).

\bibitem[{\citenamefont{Raveau et~al.}(1997)\citenamefont{Raveau, Maignan, and
  Martin}}]{Raveau97}
\bibinfo{author}{\bibfnamefont{B.}~\bibnamefont{Raveau}},
  \bibinfo{author}{\bibfnamefont{A.}~\bibnamefont{Maignan}}, \bibnamefont{and}
  \bibinfo{author}{\bibfnamefont{C.}~\bibnamefont{Martin}},
  \bibinfo{journal}{J. Solid State Chem.} \textbf{\bibinfo{volume}{130}},
  \bibinfo{pages}{162} (\bibinfo{year}{1997}).

\bibitem[{\citenamefont{Martin et~al.}(1999)\citenamefont{Martin, Maignan,
  Hervieu, and Raveau}}]{Martin99}
\bibinfo{author}{\bibfnamefont{C.}~\bibnamefont{Martin}},
  \bibinfo{author}{\bibfnamefont{A.}~\bibnamefont{Maignan}},
  \bibinfo{author}{\bibfnamefont{M.}~\bibnamefont{Hervieu}}, \bibnamefont{and}
  \bibinfo{author}{\bibfnamefont{B.}~\bibnamefont{Raveau}},
  \bibinfo{journal}{Phys. Rev. B} \textbf{\bibinfo{volume}{60}},
  \bibinfo{pages}{12191} (\bibinfo{year}{1999}).

\bibitem[{\citenamefont{Mahendiran et~al.}(2002)\citenamefont{Mahendiran,
  Maignan, Hebert, Martin, Hervieu, Raveau, Mitchell, and Schiffer}}]{PCMO}
\bibinfo{author}{\bibfnamefont{R.}~\bibnamefont{Mahendiran}},
  \bibinfo{author}{\bibfnamefont{A.}~\bibnamefont{Maignan}},
  \bibinfo{author}{\bibfnamefont{S.}~\bibnamefont{Hebert}},
  \bibinfo{author}{\bibfnamefont{C.}~\bibnamefont{Martin}},
  \bibinfo{author}{\bibfnamefont{M.}~\bibnamefont{Hervieu}},
  \bibinfo{author}{\bibfnamefont{B.}~\bibnamefont{Raveau}},
  \bibinfo{author}{\bibfnamefont{J.~F.} \bibnamefont{Mitchell}},
  \bibnamefont{and} \bibinfo{author}{\bibfnamefont{P.}~\bibnamefont{Schiffer}},
  \bibinfo{journal}{Phys. Rev. Lett} \textbf{\bibinfo{volume}{89}},
  \bibinfo{pages}{286602} (\bibinfo{year}{2002}).

\bibitem[{\citenamefont{H$\acute{e}$bert
  et~al.}(2002)\citenamefont{H$\acute{e}$bert, Maignan, Hardy, Martin, and
  Hervieu}}]{Hebert02}
\bibinfo{author}{\bibfnamefont{S.}~\bibnamefont{H$\acute{e}$bert}},
  \bibinfo{author}{\bibfnamefont{A.}~\bibnamefont{Maignan}},
  \bibinfo{author}{\bibfnamefont{V.}~\bibnamefont{Hardy}},
  \bibinfo{author}{\bibfnamefont{C.}~\bibnamefont{Martin}}, \bibnamefont{and}
  \bibinfo{author}{\bibfnamefont{M.}~\bibnamefont{Hervieu}},
  \bibinfo{journal}{Solid State Comm.} \textbf{\bibinfo{volume}{122}},
  \bibinfo{pages}{335} (\bibinfo{year}{2002}).

\bibitem[{\citenamefont{Hardy et~al.}(2003{\natexlab{a}})\citenamefont{Hardy,
  Maignan, Hebert, Yaicle, Martin, Hervieu, Lees, Rowlands, Paul, and
  Raveau}}]{Hardy03R}
\bibinfo{author}{\bibfnamefont{V.}~\bibnamefont{Hardy}},
  \bibinfo{author}{\bibfnamefont{A.}~\bibnamefont{Maignan}},
  \bibinfo{author}{\bibfnamefont{S.}~\bibnamefont{Hebert}},
  \bibinfo{author}{\bibfnamefont{C.}~\bibnamefont{Yaicle}},
  \bibinfo{author}{\bibfnamefont{C.}~\bibnamefont{Martin}},
  \bibinfo{author}{\bibfnamefont{M.}~\bibnamefont{Hervieu}},
  \bibinfo{author}{\bibfnamefont{M.}~\bibnamefont{Lees}},
  \bibinfo{author}{\bibfnamefont{G.}~\bibnamefont{Rowlands}},
  \bibinfo{author}{\bibfnamefont{D.~M.~K.} \bibnamefont{Paul}},
  \bibnamefont{and} \bibinfo{author}{\bibfnamefont{B.}~\bibnamefont{Raveau}},
  \bibinfo{journal}{Phys. Rev. B} \textbf{\bibinfo{volume}{68}},
  \bibinfo{pages}{220402(R)} (\bibinfo{year}{2003}{\natexlab{a}}).

\bibitem[{\citenamefont{Hardy et~al.}(2003{\natexlab{b}})\citenamefont{Hardy,
  Maignan, Hebert, and Martin}}]{Hardy03}
\bibinfo{author}{\bibfnamefont{V.}~\bibnamefont{Hardy}},
  \bibinfo{author}{\bibfnamefont{A.}~\bibnamefont{Maignan}},
  \bibinfo{author}{\bibfnamefont{S.}~\bibnamefont{Hebert}}, \bibnamefont{and}
  \bibinfo{author}{\bibfnamefont{C.}~\bibnamefont{Martin}},
  \bibinfo{journal}{Phys. Rev. B} \textbf{\bibinfo{volume}{67}},
  \bibinfo{pages}{024401} (\bibinfo{year}{2003}{\natexlab{b}}).

\bibitem[{\citenamefont{Carlin and van Duyneveldt}(1977)}]{Carlin}
\bibinfo{author}{\bibfnamefont{R.~L.} \bibnamefont{Carlin}} \bibnamefont{and}
  \bibinfo{author}{\bibfnamefont{A.~J.} \bibnamefont{van Duyneveldt}},
  \emph{\bibinfo{title}{Magnetic Properties of Transition Metal Compounds}}
  (\bibinfo{publisher}{Springer-Verlag}, \bibinfo{year}{1977}).

\bibitem[{\citenamefont{Radaelli et~al.}(2001)\citenamefont{Radaelli, Ibberson,
  Argyriou, Casalta, Anderson, Cheong, and Mitchell}}]{Radaelli01}
\bibinfo{author}{\bibfnamefont{P.~G.} \bibnamefont{Radaelli}},
  \bibinfo{author}{\bibfnamefont{R.~M.} \bibnamefont{Ibberson}},
  \bibinfo{author}{\bibfnamefont{D.~N.} \bibnamefont{Argyriou}},
  \bibinfo{author}{\bibfnamefont{H.}~\bibnamefont{Casalta}},
  \bibinfo{author}{\bibfnamefont{K.~H.} \bibnamefont{Anderson}},
  \bibinfo{author}{\bibfnamefont{S.-W.} \bibnamefont{Cheong}},
  \bibnamefont{and} \bibinfo{author}{\bibfnamefont{J.~F.}
  \bibnamefont{Mitchell}}, \bibinfo{journal}{Phys. Rev. B}
  \textbf{\bibinfo{volume}{63}}, \bibinfo{pages}{172419}
  (\bibinfo{year}{2001}).

\bibitem[{\citenamefont{Hardy et~al.}(2003{\natexlab{c}})\citenamefont{Hardy,
  Maignan, Martin, and Raveau}}]{Hardy03m}
\bibinfo{author}{\bibfnamefont{V.}~\bibnamefont{Hardy}},
  \bibinfo{author}{\bibfnamefont{A.}~\bibnamefont{Maignan}},
  \bibinfo{author}{\bibfnamefont{C.}~\bibnamefont{Martin}}, \bibnamefont{and}
  \bibinfo{author}{\bibfnamefont{B.}~\bibnamefont{Raveau}},
  \bibinfo{journal}{Journal of Magnetism and Magnetic Materials}
  \textbf{\bibinfo{volume}{264}}, \bibinfo{pages}{183}
  (\bibinfo{year}{2003}{\natexlab{c}}).

\bibitem[{\citenamefont{Paczuski et~al.}(1996)\citenamefont{Paczuski, Maslov,
  and Bak}}]{Bak96}
\bibinfo{author}{\bibfnamefont{M.}~\bibnamefont{Paczuski}},
  \bibinfo{author}{\bibfnamefont{S.}~\bibnamefont{Maslov}}, \bibnamefont{and}
  \bibinfo{author}{\bibfnamefont{P.}~\bibnamefont{Bak}},
  \bibinfo{journal}{Phys. Rev. E} \textbf{\bibinfo{volume}{53}},
  \bibinfo{pages}{414} (\bibinfo{year}{1996}).

\bibitem[{not()}]{note1}
\bibinfo{note}{Here we use cubic (Pm$\bar{3}$m) notation for simplicity, where
  $a=3.89\mathring{A}$. In proper orthorhombic (Pnma) notation CO occurs at
  ($\frac{1}{2}$ 0 4), AFM at ( $\frac{1}{2}$ 0 $\frac{1}{2}$) and FM at (1 0
  1) with $a=5.43594(6), b=7.66546(18), c=5.43494(14) \mathring{A}$.}

\bibitem[{\citenamefont{Lynn et~al.}(1998)\citenamefont{Lynn, Vasiliu-Doloc,
  and Subramanian}}]{Lynn98}
\bibinfo{author}{\bibfnamefont{J.~W.} \bibnamefont{Lynn}},
  \bibinfo{author}{\bibfnamefont{L.}~\bibnamefont{Vasiliu-Doloc}},
  \bibnamefont{and}
  \bibinfo{author}{\bibfnamefont{M.}~\bibnamefont{Subramanian}},
  \bibinfo{journal}{Phys. Rev. Let.} \textbf{\bibinfo{volume}{80}},
  \bibinfo{pages}{4582} (\bibinfo{year}{1998}).

\bibitem[{\citenamefont{Ghivelder et~al.}(2003)\citenamefont{Ghivelder,
  Freitas, das Virgens, Martinho, Granja, Leyva, Levy, and
  Parisi}}]{Ghivelder03}
\bibinfo{author}{\bibfnamefont{L.}~\bibnamefont{Ghivelder}},
  \bibinfo{author}{\bibfnamefont{R.~S.} \bibnamefont{Freitas}},
  \bibinfo{author}{\bibfnamefont{M.~G.} \bibnamefont{das Virgens}},
  \bibinfo{author}{\bibfnamefont{H.}~\bibnamefont{Martinho}},
  \bibinfo{author}{\bibfnamefont{L.}~\bibnamefont{Granja}},
  \bibinfo{author}{\bibfnamefont{G.}~\bibnamefont{Leyva}},
  \bibinfo{author}{\bibfnamefont{P.}~\bibnamefont{Levy}}, \bibnamefont{and}
  \bibinfo{author}{\bibfnamefont{F.}~\bibnamefont{Parisi}},
  \bibinfo{journal}{arXiv:cond-mat} \textbf{\bibinfo{volume}{0308141}},
  \bibinfo{pages}{5} (\bibinfo{year}{2003}).

\bibitem[{\citenamefont{Lynn et~al.}(1996)\citenamefont{Lynn, Erwin, Borchers,
  Huang, Santoro, Peng, and LI}}]{Lynn96}
\bibinfo{author}{\bibfnamefont{J.~W.} \bibnamefont{Lynn}},
  \bibinfo{author}{\bibfnamefont{R.~W.} \bibnamefont{Erwin}},
  \bibinfo{author}{\bibfnamefont{J.~A.} \bibnamefont{Borchers}},
  \bibinfo{author}{\bibfnamefont{Q.}~\bibnamefont{Huang}},
  \bibinfo{author}{\bibfnamefont{A.}~\bibnamefont{Santoro}},
  \bibinfo{author}{\bibfnamefont{J.-L.} \bibnamefont{Peng}}, \bibnamefont{and}
  \bibinfo{author}{\bibfnamefont{Z.~Y.} \bibnamefont{LI}},
  \bibinfo{journal}{Phys. Rev. Let.} \textbf{\bibinfo{volume}{76}},
  \bibinfo{pages}{4046} (\bibinfo{year}{1996}).

\bibitem[{\citenamefont{Dai et~al.}(2001)\citenamefont{Dai, Fernandez-Baca,
  Plummer, Tomioka, and Tokura}}]{Dai01}
\bibinfo{author}{\bibfnamefont{P.}~\bibnamefont{Dai}},
  \bibinfo{author}{\bibfnamefont{J.}~\bibnamefont{Fernandez-Baca}},
  \bibinfo{author}{\bibfnamefont{E.}~\bibnamefont{Plummer}},
  \bibinfo{author}{\bibfnamefont{Y.}~\bibnamefont{Tomioka}}, \bibnamefont{and}
  \bibinfo{author}{\bibfnamefont{Y.}~\bibnamefont{Tokura}},
  \bibinfo{journal}{Phys. Rev. B} \textbf{\bibinfo{volume}{64}}
  (\bibinfo{year}{2001}).

\bibitem[{\citenamefont{Adams et~al.}(2003)\citenamefont{Adams, Lynn,
  Smolyaninova, Biswas, Greene, II, Cheong, Mukovskii, and
  Shulyatev}}]{Adams03}
\bibinfo{author}{\bibfnamefont{C.}~\bibnamefont{Adams}},
  \bibinfo{author}{\bibfnamefont{J.}~\bibnamefont{Lynn}},
  \bibinfo{author}{\bibfnamefont{N.~V.} \bibnamefont{Smolyaninova}},
  \bibinfo{author}{\bibfnamefont{A.}~\bibnamefont{Biswas}},
  \bibinfo{author}{\bibfnamefont{R.}~\bibnamefont{Greene}},
  \bibinfo{author}{\bibfnamefont{W.~R.} \bibnamefont{II}},
  \bibinfo{author}{\bibfnamefont{S.-W.} \bibnamefont{Cheong}},
  \bibinfo{author}{\bibfnamefont{Y.}~\bibnamefont{Mukovskii}},
  \bibnamefont{and}
  \bibinfo{author}{\bibfnamefont{D.}~\bibnamefont{Shulyatev}},
  \bibinfo{journal}{arXiv:cond-mat} \textbf{\bibinfo{volume}{0304031}},
  \bibinfo{pages}{13} (\bibinfo{year}{2003}).

\bibitem[{\citenamefont{Field et~al.}(1995)\citenamefont{Field, Will, Nori, and
  Ling}}]{Field95}
\bibinfo{author}{\bibfnamefont{S.}~\bibnamefont{Field}},
  \bibinfo{author}{\bibfnamefont{J.}~\bibnamefont{Will}},
  \bibinfo{author}{\bibfnamefont{F.}~\bibnamefont{Nori}}, \bibnamefont{and}
  \bibinfo{author}{\bibfnamefont{X.}~\bibnamefont{Ling}},
  \bibinfo{journal}{Phys. Rev. Lett.} \textbf{\bibinfo{volume}{75}},
  \bibinfo{pages}{1206} (\bibinfo{year}{1995}).

\end{thebibliography}

\end{document}